\documentclass[11pt]{article}

\makeatletter
\let\NAT@parse\undefined
\makeatother

\usepackage[dvipsnames]{xcolor}

\usepackage{times}
\usepackage{graphicx}
\usepackage{fullpage}

\usepackage{hyperref}
\hypersetup{
   colorlinks,
   citecolor=Maroon,
   linkcolor=Maroon,
   urlcolor=blue
}

\usepackage{algorithm}
\usepackage{algorithmic}

\usepackage{caption}

\usepackage[noadjust, nobreak]{cite}
\usepackage{pifont}% http://ctan.org/pkg/pifont
\usepackage{comment}

\title{An Architecture for Exploiting Native User-Land Checkpoint-Restart to Improve Fuzzing}

\date{}

\author{Prashant Singh Chouhan
        \thanks{\noindent This work was partially supported by National Science Foundation Grant OAC-1740218 and a grant from Intel Corporation.}
 \\
        Northeastern University \\
        Boston, MA \\
        chouhan.p@neu.edu
\and
        Gregory Price \\
        Raytheon Technologies \\
        Annapolis Junction, MD, USA \\                              gregory.m.price@raytheon.com
\and
        Gene Cooperman$^*$ \\
        Northeastern University \\
        Boston, MA \\
        gene@ccs.neu.edu
}

\begin{document}

\maketitle
\thispagestyle{empty}
\pagestyle{plain}

%%%%%%%%%%%%%%%%%%%%%%%%%%%%%%%%%%%%%%%%%%%%%%%%%%%%%%%%%%%%%%%%%%%%%%%%%%%%%%%%
\begin{abstract}
Fuzzing is one of the most popular and widely used techniques to find vulnerabilities in any application. 
Fuzzers are fast enough, but they still spend a good portion of time to restart a crashed application and then fuzz it from the beginning.
Fuzzing an application from a point deeper in the execution is also important.
To do this, a user needs to take a snapshot of the program while fuzzing it on top of an emulator, virtual machine, or by utilizing a special kernel module to enable checkpointing.
Even with this ability, it can be difficult to attach a fuzzer after restoring a checkpoint.  As a result, most fuzzers leverage a form of fork-server design.

We propose a novel testing architecture that allows users to attach a fuzzer after the program has started running.
We do this by natively checkpointing the target application at a point of interest, and attaching the fuzzer after restoring the checkpoint.  A fork-server may even be engaged at the point of restoration.
This not only improves the throughput of the fuzzing campaign by minimizing startup time, but opens up a new way to fuzz applications.
With this architecture, a user can take a series of checkpoints at points of interest, and run parallel tests to reduce the overall state-complexity of an individual test.
Checkpoints allow us to begin fuzzing from a deeper point in the execution path, omitting prior execution from the required coverage path.
This and other checkpointing techniques are described in the paper to help improve fuzzing.

% Moreover, we can now go deeper into the execution path of the program without spending much time to get to the last checkpointed state.
% In this paper, we will explore the use of checkpointing to help improve fuzzing in many different ways.

% \textbf{distributed}

\end{abstract}
%%%%%%%%%%%%%%%%%%%%%%%%%%%%%%%%%%%%%%%%%%%%%%%%%%%%%%%%%%%%%%%%%%%%%%%%%%%%%%%%

\section{Introduction}

As software grows, we need quality, flexible tools to analyze and find bugs in the application before deployment. There are various techniques out in the market for testing.  These include SMT~\cite{barrett2008smt}, symbolic execution~\cite{baldoni2018survey}, and software fuzzing.

A widely used technique in practice is fuzzing. In this paper, we use the AFL fuzzer~\cite{afl-lop-whitepaper} (American Fuzzy Lop) for all the testing purposes. AFL is a widely adopted fuzzing framework that has been leveraged to find many bugs in large commercial applications. In practice, it has helped attackers and defenders find exploits in the software at least as commonly as other technique mentioned above. One of the reasons for its popularity is its simple architecture. The program is (optionally) recompiled with instrumentation, a set of seed inputs are provided, and AFL randomly manipulates the seeds before providing them to software under test.  The instrumentation provides AFL feedback by way of branch-coverage data.  This has made it attractive for test any software which operates on common user-input (such as file contents).

Fuzzers use various tricks and techniques to increase overall test-throughput and speed up the testing of applications. They rapidly generates different randomized inputs and execute them on the software under test, and leverage mechanisms such as fork/clone to rapidly produce new copies of the program to bypass program startup costs.

While fuzzers are fast, they still can be improved. A problem that almost all the fuzzers face is of re-initializing the target software after each successful test or after a crash.  This means a good amount of time spent in an initialization phase. If we can eliminate this start-up time, we can ultimately speed up the fuzzing process.  In addition, some tests require a multitude of interactions before user inputs are truly tested.  In some cases, it can be difficult or impossible to reproduce these interactions in a programmatic manner without significant overhead.

One of the ways to improve fuzzing is to checkpoint the target after it has been initialized.
In each test run, we can restart or fork/clone it from the checkpointed state. This not only provides a significant improvement over the conventional way of fuzzing, but we can now begin tests deeper into the execution of the program.  Additionally, by taking a series of checkpoints and fuzzing each one of them separately, we can increase the overall coverage of a program in a parallel manner, rather than spending significant time modeling specialized input/tests cases to reach that further point of execution.

Conventionally, checkpoint/restart has been done by utilizing specialized kernel modules or by running applications on top of an emulator/virtual machine.  Utilizing specialized kernel modules limits the ability of checkpointing to specific environments, kernel versions, and may or may not support fuzzing of state-full programs (such as ones which use the network).  Under emulation, one takes a complete snapshot of the virtual machine. And at each iteration, one restarts from that snapshot. Although this solves re-initialization, taking a snapshot and restoring the complete memory of the virtual machine is expensive. In addition, running an application on a virtual machine decreases the performance.

In this paper, we propose the use of DMTCP (Distributed Multi-Threaded Checkpoint), a software package for transparently checkpointing applications without modification of the target application.  This is the basis for a checkpoint-restart framework for fuzzing. In this paper AFL is used as an example, however DMTCP could be used with any fuzzer to provide quick checkpoint-restart.

Moreover, DMTCP has a plugin architecture~\cite{arya2016design}, which helps to easily modify the behavior of DMTCP. This can further help AFL in providing extra instrumentation for the target application, such as trapping on system/library function calls. By using this instrumentation, fuzzers can improve upon their feedback mechanism. Further, plugins not only provide instrumentation, they can also make clever decisions while taking checkpoints. This keeps the checkpoint logic separate from the fuzzing logic for ease of maintenance, while letting them still communicate with each other.

%%%%%%%%%%%%%%%%%%%%%%%%%%%%%%%%%%%%%%%%%%%%%%%%%%%%%%%%%%%%%%%%%%%%%%%%%%%%%%%%

\section{Background}
\label{sec:background}

In this section, we discuss the background of AFL and DMTCP.

\subsection{AFL}
\label{sec:afl}
Fuzzing is a mechanism for testing correctness of software~\cite{manes2019art}.  It can also be used for finding exploits.

AFL~\cite{afl-lop} is one of the most popular fuzzing package in use. It has had much success in general applications. It uses clever genetic algorithms to randomize the input to quickly find new interesting states in the software. AFL consists of various parts.  Here, we concentrate on:  input file; shared memory; forkserver; afl\_maybe\_log; and fuzz test child.

%\textbf{FILL IN diagram for AFL}
\begin{figure}[ht!]
  \centering
  \includegraphics[width=0.8\columnwidth]{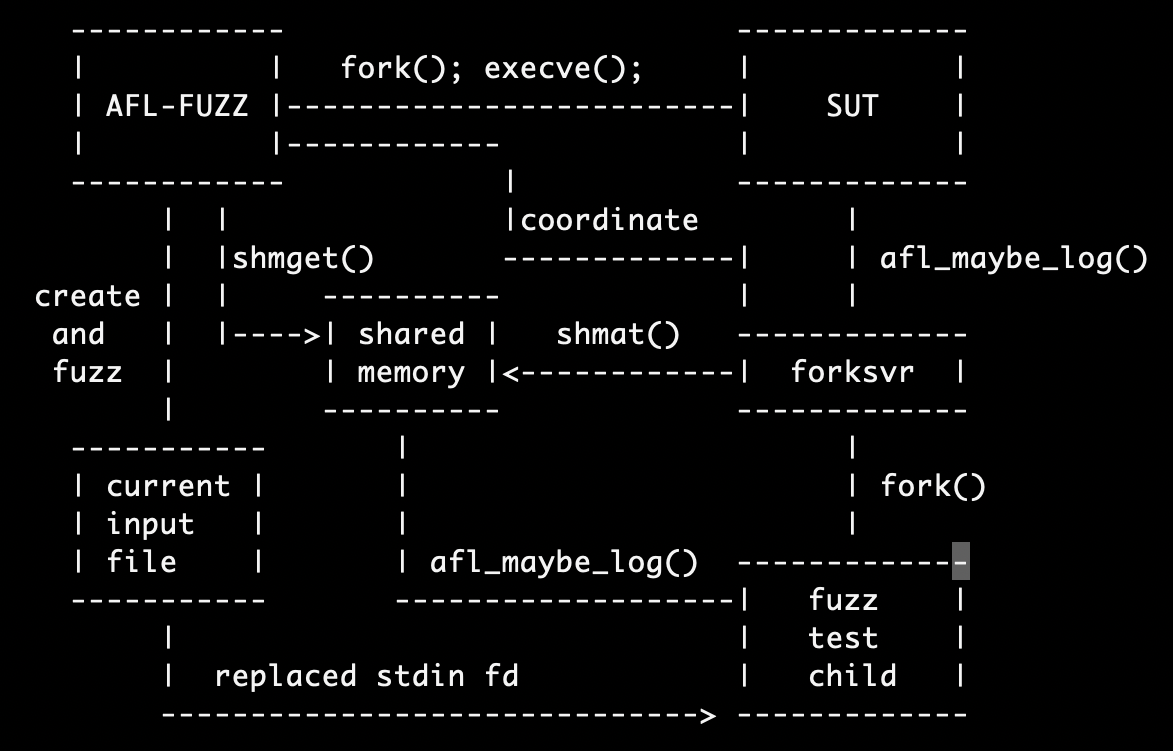}
  \caption{AFL}
  \label{fig:AFL}
\end{figure}

%\textbf{Why is a forkserver needed, what does it solve, etc.?  Also, how does the input file and the afl\_maybe\_log fit with the AFL design? Also, what about the role of shared memory for communication between the ``fuzz test child'' and AFL?}

\paragraph{Input file} 
In a trivial setup, AFL replaces the stdin of the target program with a file, known as the input file. So, instead of reading input from stdin, the target program reads the input from the input file. When the program finishes executing, AFL modifies the contents of the input file before beginning a new test. This pattern is followed throughout the complete fuzzing process.

\paragraph{afl\_maybe\_log}
This is a function that contains the implementation of forkserver and all the logic for recording the code coverage. It is injected during recompilation after the call to main() and before each branch instruction. On first invocation, it initializes the forkserver or disable instrumentation if a forkserver is not found (this enables the user to run the program without AFL attached). It then communicates with the fuzzer about the status of the forked child. On each branch hit, it updates the shared memory. 

\paragraph{Forkserver}
In a basic approach for fuzzing, a fuzzer would fork and exec the target application. This approach is simple, but it has a high overhead. Each exec() calls incurs significant linking and library initialization overhead. To alleviate this, AFL utilizes a forkserver design. On the first call to afl\_maybe\_log(), the application creates a communication pipe with AFL. When AFL sends a ``go'' command to initiate a test, the instrumentation forks the program with copy-on-write, which is very quick and lightweight (some newer AFL version may allow the use of clone() to reduce even more overhead). The forkserver (parent application) now calls waitpid() on the child and sends control information back to AFL via the communication pipes. This complete setup helps reduce the high setup overhead time.  A more complex design is required to push distributed fuzzing jobs to large server environments.

\paragraph{Shared memory}
AFL uses a shared memory buffer to collect branch coverage of the target application during tests. The forkserver is responsible for mapping this buffer, and children inherit a copy of this mapping for use with each subsequent afl\_maybe\_log() call. After each execution of the "fuzz test child", AFL reads the shared memory buffer and tweaks the input file accordingly. The shared memory is then reset before the next test.  

%It speeds up the process of forking the children by eliminating a lot of initial setup required by fork() followed by exec(). It is basically a small piece of code being injected into the application itself, so it is the same as the application. When the program first hits afl\_maybe\_log, it gets set up and forms all the connections to the fuzzer and the shared memory. Then it keeps on forking itself. 
\paragraph{Fuzz Test Child}
The fuzz test child is the forked child of the forkserver. ``fuzz test child'' executes from where the program left of (main() in the trivial setup), and all the branches executed by it are logged by afl\_maybe\_log.  afl\_maybe\_log updates the shared memory.  The shared memory is read by AFL, to see the coverage.  Based on the coverage input, AFL modifies the current input file for the next ``fuzz test child''.  The program will utilize the input file as its stdin for all future reads from stdin, this is what enables fuzzing the program.

\subsection{Distributed Multi-Threaded Checkpoint (DMTCP)}
\label{sec:dmtcp}

DMTCP~\cite{ansel2009dmtcp} is a software package that has traditionally been used to checkpoint and restart HPC applications~\cite{cao2016system}.  DMTCP can transparently checkpoint applications, with no modification to the target application or the operating system.  It works for centralized and distributed computations.  It does not require any special user privileges.  It operates entirely within user space.  It can be used with MPI, Matlab, Python, Perl, and a series of other packages.

DMTCP also includes a plugin architecture~\cite{arya2016design}.  DMTCP provides a plugin architecture which can be used in two ways.  First, it provides event hooks on the events provided by DMTCP.  Second, it provides wrappers around functions via the LD\_PRELOAD functionality.  A DMTCP plugin is in the form of a shared library.  So, it can be dynamically loaded.

\begin{figure}[ht!]
  \centering
  \includegraphics[width=0.5\columnwidth]{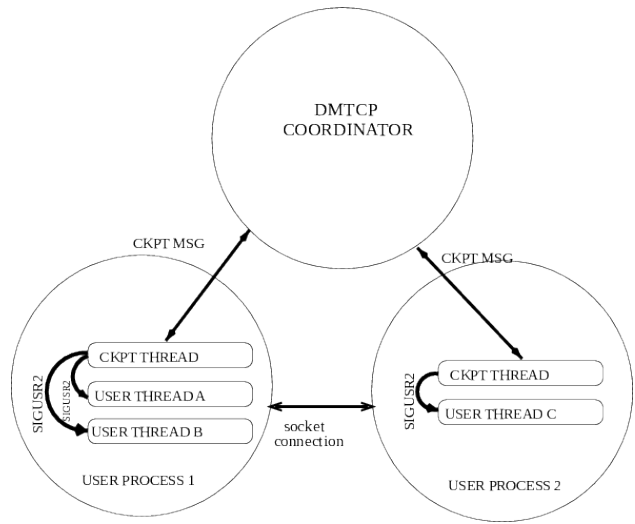}
  \caption{Architecture of DMTCP}
  \label{fig:dmtcp}
\end{figure}

Using DMTCP allows us to wrap or replace library and system calls.  This can be used to provides extra instrumentation or direct the usage of fuzzed input to certain function calls.  In this work, we use this for detecting patterns of system calls or library calls by the applications.  Wrappers are placed around read, write, and seek in the examples in this work.

DMTCP also provides event hooks. Using the event hooks, we can do additional or specialized setup during testing.  For example, we can reset kernel state (replaying a set of system calls) or reset global variables before restart so that they appear the same as prior to checkpoint.

Using event hooks and function wrappers together in DMTCP provides \emph{process virtualization}~\cite{arya2016design}. Process virtualization is needed
after the program has been restarted when certain system resources will differ (for example, each child will have different process ids).
Wrapper functions and event hooks can (and are) used to pass a virtual process id to the target application instead of the real process id.
  The application might have cached the earlier process ids.  It may want to use those cached process ids again, for example to send a signal.  DMTCP hides this by creating wrappers around all library or system calls that use a process id.  See~\cite{arya2016design} for details.

This functionality can be extended to wrap any dynamically loaded library via the plugin system.  For example, it could be utilized to recreate a specific network state should the program require an open session to resume from that point in time.
%used to hide arbitrary process ids allotted by kernel from the restarted application. Restarted application just sees the process ids it had received during the initial run.

%\textbf{Describe:  afl\_maybe\_log,  AFL forkserver.}

\section{Design}

In this section, we describe the technical design details of how to marry AFL and DMTCP together. 
We do this in three steps:
\begin{enumerate}
    \item We first run the (instrumented) application without AFL under dmtcp\_launch.
    \item When the program reaches the desired state, we request a checkpoint.
    \item Then we restart the process with AFL attached.
\end{enumerate}
 %*** This displays how we can attach AFL later in time.

%Since we have already talked about AFL and DMTCP before, we marry them together next.

AFL will only be used during the third phase.  During the first two phases, the program can interacted with and DMTCP is free to collect administrative data (such as network or file session details) that are needed to successfully restart a checkpoint.

\paragraph{Launching under DMTCP}

\begin{figure}[ht!]
  \centering
  \includegraphics[width=0.8\columnwidth]{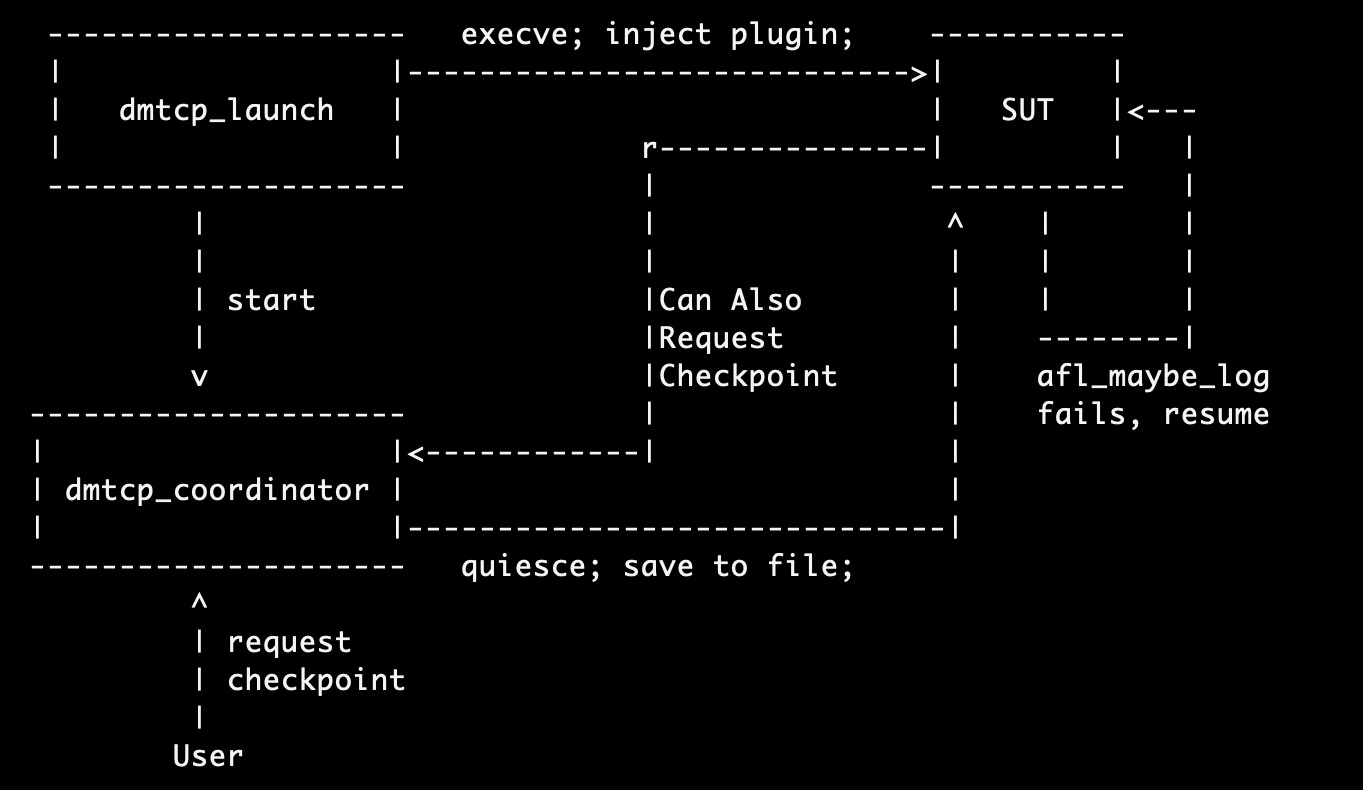}
  \caption{Launch under DMTCP}
  \label{fig:run}
\end{figure}

Diagram~\ref{fig:run} describes how an application is launched under DMTCP.
The target application is compiled using afl-gcc. afl-gcc is a wrapper around gcc, which basically inserts the fork server and branch coverage instrumentation and the fork server.  The target application is then run under dmtcp\_launch.  After executing the command, dmtcp\_launch connects with the DMTCP coordinator.  A DMTCP coordinator is created by dmtcp\_launch if one is not already there.  The DMTCP coordinator is used to enable external entities to control the checkpointing of a program.

Then dmtcp\_launch does some setup and execs into the target application:  System Under Test (SUT).  Since we compiled the application SUT with afl-gcc, it contains instrumented branches.  On the first instrumented branch, afl\_maybe\_log tries to run the AFL forkserver.  If there is any kind of failure (in this case, AFL is not available), then afl\_maybe\_log sets a failure flag to ABORT, and aborts any further setup.  Subsequent calls to afl\_maybe\_log do nothing and they just return.  The program continues executing as-if it wasn't instrumented.

The application can optionally make calls to dmtcp\_checkpoint() to request a checkpoint.  For our testing purposes, this call has been added directly to a target application, although the checkpoint can also be triggered externally.  This sends a request to the DMTCP coordinator.  The coordinator checkpoints the application and the checkpoint image is saved to a file.

Notably, to support AFL, we must launch DMTCP and the application with an AFL-aware DMTCP plugin, which virtualizes the effects of afl\_maybe\_log() and makes it possible to attempt re-initialization after restarting from a checkpoint.

\paragraph{Restarting with AFL attached}

\begin{figure}[ht!]
  \centering
  \includegraphics[width=0.8\columnwidth]{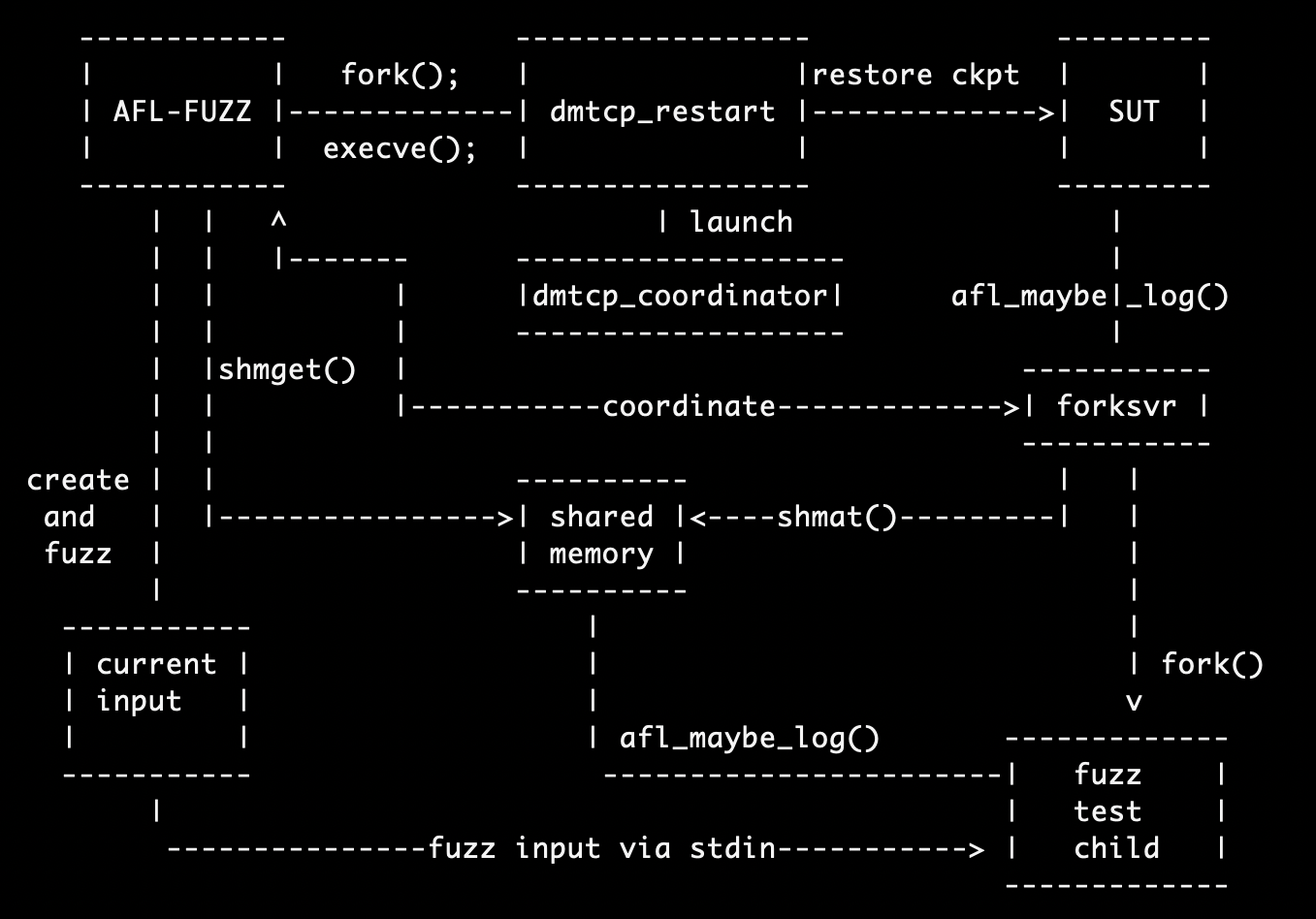}
  \caption{Restart with AFL}
  \label{fig:restart}
\end{figure}

Diagram~\ref{fig:restart} describes how an application is restarted with AFL attached.
At the time of restart, we run AFL.  However, instead of running AFL directly on the application, we substitute this for the checkpointed application using dmtcp\_restart:\hfil\break
\hbox{\ \ \ \ } {\tt dmtcp\_restart ckpt\_image.dmtcp}\hfil\break
where {\tt ckpt\_image.dmtcp} is the checkpoint image that was saved.  AFL first makes a call to shm\_get(), and sets up the shared memory.  Then, AFL forks and execs into the dmtcp\_restart command.  

Next, dmtcp\_restart launches the DMTCP coordinator, and restores (restarts) the checkpointed process. The checkpoint image being restored by dmtcp\_restart contains the special AFL plugin (from the command line of dmtcp\_launch), which resets the afl\_maybe\_log() failure flag back to zero (its initial state).

Finally, dmctp\_restart passes the control to SUT (the target application). SUT contains multiple calls to afl\_maybe\_log. The next call to afl\_maybe\_log starts the forkserver, which sets up the shared memory, and connects to AFL.  After all of this is done, it forks a new ``fuzz test child'', and monitors its return status to determine if the process exits normally or if it crashes.  The forkserver passes this information back to AFL, so that the fuzzer is aware of the status of the process being fuzzed.

\subsection{Strategy for DMTCP plugins to control the instrumented binary}
\label{sec:strategy}

The plugin architecture of DMTCP can be used provide more interesting forms of instrumentation without directly modifying the system under test.  This helps us get a better understanding of the program, and we can propagate this information to the fuzzer. Based on this, the fuzzer can generate a more informed input.  Using the plugin architecture of DMTCP,
we can support checkpoint fuzzing by implementing it separately from the mutation engine (in this case, AFL).  This separation of concerns is important for both development and for maintainability.

In the current design, the SUT must be compiled into an instrumented binary.
This involves injecting the AFL forkserver code into the binary, and instrumenting all the branch instructions of the binary to monitor the code coverage.

The compiled binary is then run under dmtcp\_launch, and without AFL attached.  Without the fuzzer, the binary runs normally as if there was no instrumentation.  We then call dmtcp\_checkpoint to checkpoint the binary.  

We restart the checkpointed binary under AFL-fuzz, using dmtcp\_restart.
% At the time of restart, we first run AFL-fuzz and pass it the dmtcp\_restart binary and the checkpointed file.
dmtcp\_restart includes a plugin that resets the forkserver state, and attempts to re-initialize the fork-server.

Next, our restarted binary operates as if DMTCP isn't there.  The logging and coverage information is sent back to AFL from the restarted binary, as if it were running normally without any restart.

This shows that the restarted application can run successfully under dmtcp\_restart with AFL attached.  This opens up many possibilities, including the example of the next section.

\subsection{Example:  Three plugins to monitor the system under test}

Here, we propose how three custom plugins can be created for monitoring a specific system under test.
These plugins checkpoint the application, analyze it, and reset the restarted application.

We first describe a plugin for checkpointing.  As we have described in the DMTCP plugin architecture of Section~\ref{sec:dmtcp}.  We can checkpoint the application according to a predefined pattern, and we can express that pattern in the context of a DMTCP plugin.  For example, consider an application with a pattern of initializing itself with 5 calls to read() which are known to contain no bugs.  By checkpointing after the first 5 reads, this eliminates this overhead.  This does not require modifying the original binary, so it can even be used with dumb-fuzzing mode (no instrumentation).

We next describe an analysis plugin.  An analysis plugin is would hook a variety of system and library calls to determine when a the program produces a previously unseen execution pattern.  The plugin can then checkpoint at each function call, providing a new system state from which to begin fuzzing from.

Last, we describe a reset plugin. A reset plugin is a plugin used to reset the AFL forkserver, and resets any resources (such as file descriptors or network connections) that are needed at the time of attaching AFL to the restarted process.

\subsection{Implications:  a distributed fuzzer tree}

% If we promise three plugins, we will need at least three sentences --- one for each plugin.
Here, we describe how the DMTCP plugins can potentially be used to build a distributed fuzzer tree. First we start fuzzing with a random input seed. Then on each newly discovered pattern of library/system calls, we create a checkpoint using the checkpoint plugin. Each checkpoint image file can be restarted and fuzzed independently.  Each checkpoint image represents a distinct node in an ``execution state tree''.

We continue to take checkpoints based on newly detected patterns, using the analysis and checkpoint plugins. And each time we restart, the reset plugin resets the forkserver code in the binary to the appropriate state.

Importantly, since DMTCP can transparently virtualize resources (such as file descriptors and network sessions), fuzzing campaign developers needn't worry about writing custom code to handle those cached resources.  This would make automated execution state exploration significantly easier.

\section{Related work}

AFL-QEMU~\cite{afl-qemu} is an extension of the original AFL fuzzer~\cite{afl-lop-whitepaper} to do black-box fuzzing of closed-source binaries. It runs QEMU in user-mode to minimize the overhead of running an application over a virtual machine. User-mode QEMU just emulates a CPU and translates all the system calls to the native OS syscalls and avoids the overhead of emulating all the h/w resources and a full kernel. QEMU has built-in support to take a snapshot of the complete virtualized machine, and restore it later in time. Even after using QEMU with user-mode, it has significantly higher overhead than a native execution of the target program. 

Another work employs LLVM-based instrumentation so that AFL-fuzz provides compiler-based instrumentation for compiling it into assembly~\cite{afl-llvm}.  This helps optimize the code that is used at runtime.  It also makes the instrumentation CPU-independent.

For our purposes, the most interesting application of LLVM-based AFL-fuzz is that we can attach AFL after AFL-init. This helps in resetting the forkserver, and running from there on.

Another approach to improving fuzzing is to use in-memory checkpoints~\cite{xu2017designing}. The fuzz time is improved by taking a snapshot of the process and replicating it, instead of calling fork.  This is similar to a checkpoint in terms of reading the complete memory and saving it, but it does it in RAM instead of saving to secondary storage.  This improves the fuzz time of the application.

% \textbf{
% We'll refer to these citations~\cite{afl-lop,afl-qemu,afl-llvm,xu2017designing}
% }

\section{Conclusion}

As described in the paper, one can easily checkpoint and restart an application after some initialization setup and still dependably execute library and system calls with virtualized resources.  The software architecture presented here would improve the performance of the fuzzer by some constant, whose magnitude is to be determined in a future implementation.  Also, it was shown that DMTCP's plugin architecture can help extend coverage capabilities of fuzzer, and thus help improve the code coverage or state exploration, while decreasing the overall execution time.  DMTCP's distributed architecture can also allow us to extend this approach to distributed fuzzers in future work.

Hence, the new architecture provides a simple, yet powerful tool, which can transparently apply checkpoint fuzzing to target applications, once those applications have been compiled using afl-gcc.

\bibliographystyle{IEEEtran}
\bibliography{fuzzing}

\begin{thebibliography}{10}
\providecommand{\url}[1]{#1}
\csname url@samestyle\endcsname
\providecommand{\newblock}{\relax}
\providecommand{\bibinfo}[2]{#2}
\providecommand{\BIBentrySTDinterwordspacing}{\spaceskip=0pt\relax}
\providecommand{\BIBentryALTinterwordstretchfactor}{4}
\providecommand{\BIBentryALTinterwordspacing}{\spaceskip=\fontdimen2\font plus
\BIBentryALTinterwordstretchfactor\fontdimen3\font minus
  \fontdimen4\font\relax}
\providecommand{\BIBforeignlanguage}[2]{{%
\expandafter\ifx\csname l@#1\endcsname\relax
\typeout{** WARNING: IEEEtran.bst: No hyphenation pattern has been}%
\typeout{** loaded for the language `#1'. Using the pattern for}%
\typeout{** the default language instead.}%
\else
\language=\csname l@#1\endcsname
\fi
#2}}
\providecommand{\BIBdecl}{\relax}
\BIBdecl

\bibitem{barrett2008smt}
C.~Barrett, ``{SMT} solvers: Theory and practice,'' in \emph{Summer School
  2008: Verification Technology, Systems \& Applications}, 2008,
  \url{https://resources.mpi-inf.mpg.de/departments/rg1/conferences/vtsa08/slides/barret2_smt.pdf}.

\bibitem{baldoni2018survey}
R.~Baldoni, E.~Coppa, D.~C. D’elia, C.~Demetrescu, and I.~Finocchi, ``A
  survey of symbolic execution techniques,'' \emph{ACM Computing Surveys
  (CSUR)}, vol.~51, no.~3, pp. 1--39, 2018, also see:
  \url{https://arxiv.org/pdf/1610.00502.pdf}.

\bibitem{afl-lop-whitepaper}
M.~Zalewski, ``Technical whitepaper for afl-fuzz,''
  \url{https://lcamtuf.coredump.cx/afl/technical_details.txt}; also see:
  \url{https://lcamtuf.coredump.cx/afl/}.

\bibitem{arya2016design}
K.~Arya, R.~Garg, A.~Y. Polyakov, and G.~Cooperman, ``Design and implementation
  for checkpointing of distributed resources using process-level
  virtualization,'' in \emph{IEEE Int. Conf. on Cluster Computing
  (CLUSTER'16)}.\hskip 1em plus 0.5em minus 0.4em\relax IEEE Press, 2016, pp.
  402--412.

\bibitem{manes2019art}
V.~J.~M. Man{\`e}s, H.~Han, C.~Han, S.~K. Cha, M.~Egele, E.~J. Schwartz, and
  M.~Woo, ``The art, science, and engineering of fuzzing: A survey,''
  \emph{IEEE Transactions on Software Engineering}, 2019.

\bibitem{afl-lop}
M.~Zalewski, ``American fuzzy lop (2.52b),''
  \url{https://lcamtuf.coredump.cx/afl/}.

\bibitem{ansel2009dmtcp}
J.~Ansel, K.~Arya, and G.~Cooperman, ``{DMTCP}: Transparent checkpointing for
  cluster computations and the desktop,'' in \emph{2009 IEEE International
  Symposium on Parallel \& Distributed Processing (IPDPS'09)}.\hskip 1em plus
  0.5em minus 0.4em\relax Rome, Italy: IEEE, 2009, pp. 1--12.

\bibitem{cao2016system}
J.~Cao, K.~Arya, R.~Garg, S.~Matott, D.~K. Panda, H.~Subramoni, J.~Vienne, and
  G.~Cooperman, ``System-level scalable checkpoint-restart for petascale
  computing,'' in \emph{22nd IEEE Int. Conf. on Parallel and Distributed
  Systems (ICPADS'16)}.\hskip 1em plus 0.5em minus 0.4em\relax IEEE Press,
  2016, pp. 932--941, also, technical report available as: arXiv preprint
  arXiv:1607.07995.

\bibitem{afl-qemu}
M.~Zalewski \emph{et~al.}, ``google/{AFL}/qemu\_mode (github),''
  \url{https://github.com/google/AFL/tree/master/qemu_mode}.

\bibitem{afl-llvm}
L.~Szekeres \emph{et~al.}, ``Fast {LLVM}-based instrumentation for afl-fuzz,''
  \url{https://github.com/google/AFL/tree/master/llvm_mode}.

\bibitem{xu2017designing}
W.~Xu, S.~Kashyap, C.~Min, and T.~Kim, ``Designing new operating primitives to
  improve fuzzing performance,'' in \emph{Proceedings of the 2017 ACM SIGSAC
  Conference on Computer and Communications Security}, 2017, pp. 2313--2328,
  \url{https://gts3.org/assets/papers/2017/xu:os-fuzz.pdf}.

\end{thebibliography}

\end{document}